\documentclass[12pt]{article}

\usepackage[dvips]{graphicx}
\usepackage{amssymb,amsfonts,amsmath}
\usepackage{authblk}

\begin{document}

\title{Financial crises and the evaporation of trust}

\author[,1]{Kartik Anand\footnote{To whom this correspondence should be addressed. Email: kanand@ictp.it}}
\author[,2]{Prasanna Gai}
\author[1]{Matteo Marsili}
\affil[1]{The Abdus Salam International Centre for Theoretical Physics, Strada Costiera 11, 34014 Trieste, Italy.}
\affil[2]{The Australian National University, Crawford School of Economics and Government, Canberra ACT 0200, Australia.}

\maketitle

\begin{abstract}
Trust lies at the crux of most economic transactions, with credit markets being a notable example.  Drawing on insights from the literature on coordination games and network growth, we develop a simple model to clarify how trust breaks down in financial systems.  We show how the arrival of bad news about a financial agent can lead others to lose confidence in it and how this, in turn, can spread across the entire system.  Our results emphasize the role of hysteresis -- it takes considerable effort to regain trust once it has been broken.  Although simple, the model provides a plausible account of the credit freeze that followed the global financial crisis of 2007/8, both in terms of the sequence of events and the measures taken (and being proposed) by the authorities.
\end{abstract}

\section{Introduction}
Trust is ubiquitous in social and economic activity, irrespective of underlying cultural differences. Drawing from the literature on social capital \cite{Glaeser2000}, trust is defined as ``the commitment of resources to an activity where the outcome depends upon the cooperative behavior of others". Moreover, it seems to obey a type of self-reinforcing dynamic \cite{Bagehot1873,Hosking2008} -- individuals continue to trust beyond the point where evidence points to the contrary.  Eventually, however, the accumulated weight of evidence turns them towards distrust, which is equally reinforcing.   

Credit markets exemplify the way in which trust lies at the very foundation of modern financial systems. Indeed, the very word ``credit" derives from the Latin, {\it credere}, which means to trust. If credit is readily available, it enables us to derive goods and services from those whom we do not know or have any reason to trust. Schumpeter \cite{Schumpeter1939} highlights the critical role of credit in real economic activity, noting that ``capitalism is that form of private property economy in which innovations are carried out by means of borrowed money [credit]."

Viewed in this light, a proximate cause of the global financial crises of 2007/08 has been the generalized breakdown of trust between banks and investors in banks. The triggers were revelations of losses on United States sub-prime mortgages and  other toxic financial assets by banks. An immediate consequence was a freeze in interbank money markets, as banks ceased lending to each other. Mounting funding pressures, in turn, lead to questions about banks' future profitability and, in some cases, viability.

\begin{figure}[h]
\begin{center}
\includegraphics[width=13cm]{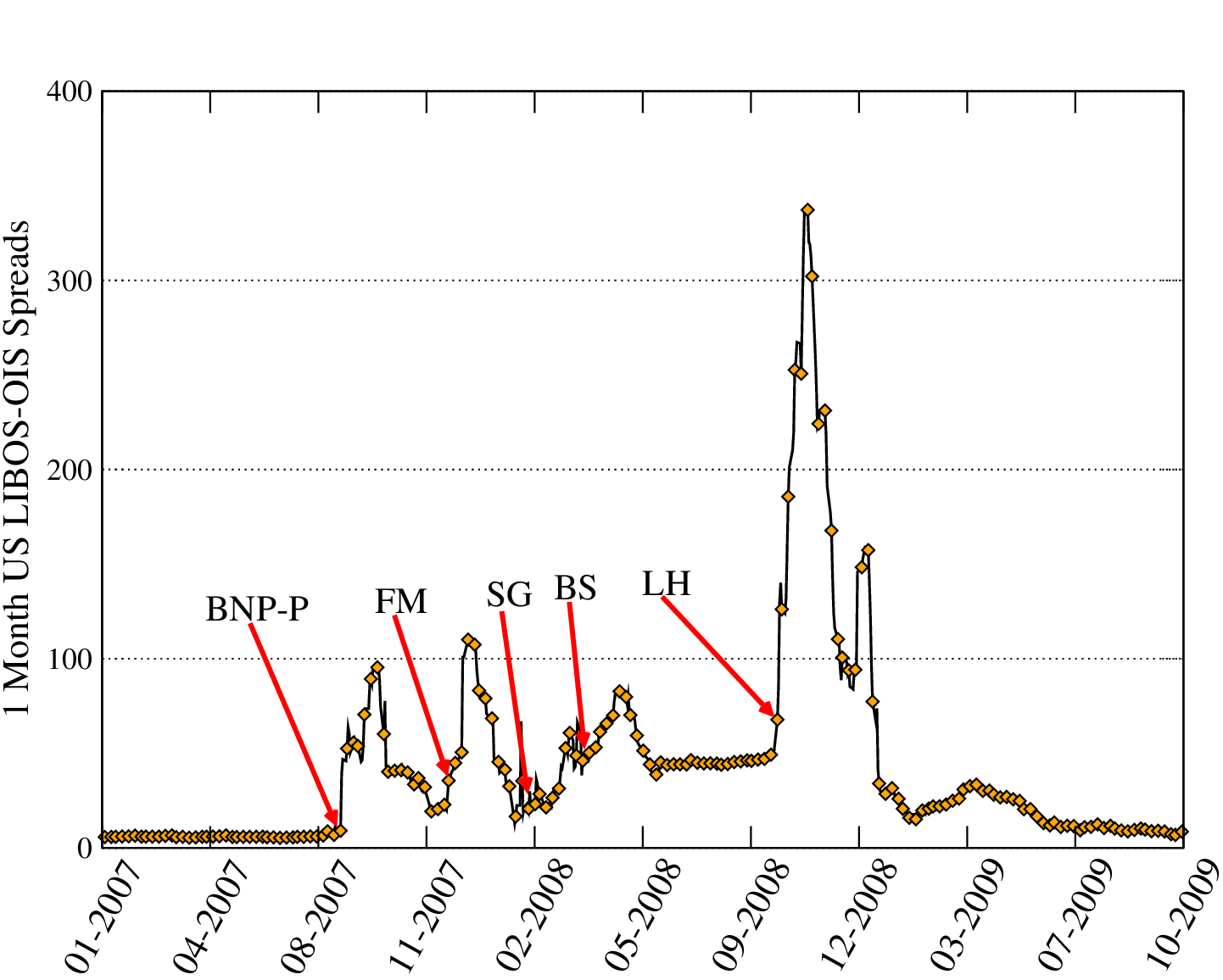}
\end{center}
\caption{1 Month US LIBOR-OIS Rates, in basis points, during the financial crisis of 2007-08. The events highlighted are: (i) August 9, 2007 -- BNP Paribas suspends calculation of asset values of three money market funds exposed to US sub-prime mortgages; (ii) November 20, 2007 -- Freddie Mac announces losses for the third quarter of 2007; (iii)  January 24, 2008 -- Soci\'et\'e G\'en\'erale reveals trading losses resulting from fraudulent activities by a single trader; (iv) March 13, 2008 -- Bear Stearns files for bankruptcy, and (v) September 15, 2008 -- Lehman Brothers files for bankruptcy.} \label{fig-libor}
\end{figure}

Fig. \ref{fig-libor} illustrates how the arrival of news of losses at troubled hedge funds, downgrades of structured financial products, and concerns about asset quality increased funding pressures on banks.  Before the crisis, banks required some 10 basis points of compensation for making one month loans to each other.  By September 2007, that compensation premium had risen to around 100 basis points.  The ensuing collapse of the investment banks Bear Sterns and Lehman Brothers in 2008 led the premium to rise more than thirty-fold from pre-crisis levels. And, despite public sector bailouts of the banking system in the major economies, trust has been slow to return.

While specific details and rigorous economic modelling are necessary to properly understand the financial crisis, there is nonetheless scope to clarify the mechanisms by which trust evaporates in networks in general, and financial networks in particular. When credit markets break down, {\em strategic uncertainty} -- i.e., uncertainty about the actions of other participants -- can be more important than {\em structural uncertainty} -- i.e., uncertainty concerning the soundness of fundamentals \cite{MorrisShin2003,MorrisShin2008}. In the case of a bank run, for example, depositors may consider withdrawing their funds if they believe that other depositors are also going to withdraw their savings. Such behavior can be triggered by doubts over the banks' balance sheet position (structural uncertainty), or by fears that others may withdraw their funds (strategic uncertainty). With imperfect common knowledge of fundamentals, the arrival of bad news about fundamentals leads small seeds of doubt to reverberate across all lenders, leading potentially to a wholesale withdrawal of lending and the bankruptcy of the counterparty.

Morris and Shin \cite{MorrisShin2003,MorrisShin2008} formalize such situations as a coordination game between lenders involved with a single, risky, counter-party. Their analysis does not, however, address the issue of wide-spread contagion at the system level. As the recent crisis makes clear, small seeds of doubt about one counter-party reverberated across the entire global financial system, enveloping credit markets from New York to Sydney. Agents are likely to be involved in as many coordination games as the number of agents they are lending to.  In addition, as borrowers, they are also party to coordination games being played by the agents lending to them. In short, in realistic financial market settings, many coordination games take place simultaneously, with players having less than full information about the balance sheet positions of their counterparties.

In this paper, we extend the insight offered by coordination games to the system level using a model of network growth.  Specifically, we show how the arrival of signals about counterparties sows the seeds of distrust and triggers foreclosures across a large population. The model, thus, helps shed light on the recent ``freeze" in global interbank lending, in which banks ceased lending to each other, including to well known and long-standing counterparties, once negative signals began to accumulate.

Our results may be summarized as follows: the financial system can converge to a ``good" equilibrium in which a dense network of credit relations exists and the risk of a run, and subsequent default, is negligible.  But a ``bad" equilibrium is also possible -- here the credit network is sparse because investors are more skittish and prone to prematurely foreclosing their credit relationships.  The transition between the two equilibria is sharp and both states exhibit a degree of resilience; once a crisis tips the system into the sparse state, the restoration of trust requires considerable effort, with model parameters needing to shift well beyond the turning point. And when the system reverts to a good state, it is robust even to deteriorating conditions.

A crucial feature of our model is the rate at which bad news about the creditworthiness of an agent arrives.  This, together with the maturity structure of debt contracts, determines the (endogenous) rate of link decay in the network.  Intuitively, when bad news arrives an agent may be forced into default by the ensuing foreclosures. This leads to a rearrangement of balance sheets across the financial system -- agents who have lent to it loose assets, while agents who borrowed from the defaulter loose liabilities. As a result, there is a possibility that some counterparties may be placed under stress, precipitating further rounds of foreclosures. We discuss the properties of the stationary state of these processes.

Our paper complements recent work on interbank contagion.  On the empirical front, researchers have relied on counterfactual simulations based on available interbank exposure data to estimate the probability and spread of contagion (see \cite{Upper2007} for a review).  The key finding is that network contagion is unlikely but, if it were to take place, can lead to the breakdown of a substantial fraction of the banking system.  Unlike our paper, these analyses treat the underlying topology of interactions and the balance sheets of the agents as static.

Theoretical literature in the area builds upon Allen and Gale \cite{Allen2000} and focuses on optimal behavior in small networks. For example, Caballero and Simsek \cite{Caballero2009} perturb a financial network in order to model the spread of contagion. They appeal to the rising costs of understanding the structure of the network as the basis for complexity. If information about the network structure is costless, there is no foreclosure. But if, following a shock, these information costs rise sharply, banks' inability to understand the structure of the (small) network to which they belong leads them to withdraw from their loan commitments.

In a related contribution, Acharya et. al. \cite{Acharya2009} also highlight the role played by information arrival relative to rollover frequency.  While their model is also about runs in credit markets, they do not adopt a network approach.  Moreover, their focus is on the debt capacity of assets used as collateral for short-term borrowing, whereas our emphasis is on the self-reinforcing dynamics of trust in the face of information disclosure.

The remainder of this paper in organized as follows. In Sec. \ref{sec:model} we begin by presenting a solution to the single instance coordination game using the global games framework. This is followed by our network growth model for credit markets. In Sec. \ref{sec:results} we provide simulation and numerical results for our model, followed by a simple analytical characterization for the stationary states. We conclude in Sec. \ref{sec:discussion} by describing how our model provides insight to the financial crises of 2007/08 and add to the evolving regulatory policy discussion. We relegate some of the more technical details to the appendices. 

\section{The model}
\label{sec:model}
Consider a population of $N$ agents engaged in bilateral credit relationships with each other.  A financial system of this kind can be viewed as a directed network, with nodes representing the agents and outgoing links reflecting loans from one agent to another.  To keep matters simple, suppose that all loans take the same nominal value.

The financial position of agent $i$ is summarized by the assets and liabilities on its balance sheet.  Assets include holdings of cash, $b_i^0$, as well as loans made to other agents, $b_i$.  Liabilities, namely the monies owed by agent $i$ to its counterparties, are denoted by $\ell_i$ and reflect (the number of) incoming links. Since every liability is someone else's asset, every outgoing link for one node is an incoming link for another node. So the total amount of assets in the system matches the total liabilities or, equivalently, the average in-degree equals the average out-degree $\langle b \rangle = \langle \ell \rangle$, where the angled bracket refers to the average over all agents. That said, individual agents may be in surplus or deficit in their individual financial positions.  The average connectivity, $\rho = \langle \ell \rangle$ of the network offers a summary measure of the extent of global financial market integration in what follows.

The credit network is dynamic, with debt contracts (or links) continuously being established and terminated as they reach maturity. The dynamic evolution of the network is punctuated by episodes where the $\ell_i$ lenders of agent $i$ engage in a game to decide whether to prematurely foreclose their loans to $i$. We first describe this foreclosure game, before clarifying the dynamics of the network.

\subsection{Foreclosure game}

Imagine that, at a particular time $t_\nu$, agent $i$ has $\ell_i$ liabilities, $b_i$ assets and $b_i^0$ amount of cash. This information is disclosed to all $\ell_i$ lenders, who are then given the choice of withdrawing their funds (foreclosing) or rolling them over to maturity. We follow the analysis of Morris and Shin \cite{MorrisShin2008} in describing this situation. Accordingly, their paper and Appx. \ref{apx:foreclose} provide a detailed game-theoretic account, while we limit the discussion below to the key elements.

For each creditor $j$, foreclosure yields a payoff of zero, whereas rolling over yields a payoff of $1-c_j>0$, provided that the number of lenders who opt out does not exceed $b_i+b_i^0$, on the asset side of agent $i$'s balance sheet. If, however, more than $b_i+b_i^0$ agents opt out, this depletes the financial resources of agent $i$, who is forced into default. This results in lender $j$, who decided to roll over, to incur a loss of $c_j$. Following \cite{MorrisShin2008}, we refer to $c_j$ as the {\it cost of miscoordination} to $j$. Intuitively, when the cost of miscoordination is high, coordination is more difficult to achieve since the opportunity cost of remaining in the investment is greater. The payoff matrix for agent $j$, in terms of the number $\ell_i'$ of lenders who foreclose, is therefore

\begin{equation}
\label{payoff-mat}
\begin{array}{c|c|c} & \ell_i'\le b_i+b_i^0 & \ell_i'> b_i+b_i^0 \\\hline \hbox{foreclose} & 0 & 0 \\\hline \hbox{roll~over} & 1-c_j & -c_j\end{array}\,.
\end{equation}

If no agent were to foreclose, it would be convenient to roll over for all of them, as long as $c_j<1$. But if an agent nurture doubts on others foreclosing, this may prompt the agent to consequently foreclose, making those doubts self-fulfilling. Agents know their cost $c_j$ and that the costs of other agents are drawn from some distribution. In this manner, one breaks the assumption of common knowledge. The ambiguity faced by an agent, in not knowing the costs faced by others,  plants the seed of doubt for strategic uncertainty.

The unique Nash equilibrium for this game \cite{MorrisShin2008} amounts to choosing a ``switching strategy" for each counterparty, $j$:
\begin{equation}
\label{cj}
\left\{\begin{array}{lr}\hbox{rollover} &\hbox{if}\quad c_j\le c^{\star}
\\
\hbox{foreclose} &\hbox{if}\quad c_j> c^{\star}\end{array}
\right.\,,
\quad \hbox{with}
\quad c^{\star}\equiv \frac{b_i+b_i^0+1}{\ell_i+1}.
\end{equation}
We now provide the basic intuition for this solution.

Assume that all counter-parties are subject to switching strategies, i.e., $j$ will rollover its' loan, if $c_j\le c^{\star}$, or foreclose, otherwise. We wish to define $c^{\star}$ in terms of $i$'s balance sheet. To this end, we introduce the probability $\phi$ that no more than $b_i+b_i^0$ agents have cost greater than $c^{\star}$ -- and hence foreclose their loans to $i$. In order to compute $c^{\star}$, consider the position of an agent $h$ with $c_h=c^{\star}$. This implies that $h$ must be indifferent between foreclosing and rolling over. On equating the expected payoffs corresponding to the two options, we get $\phi=c^{\star}$. Furthermore, $\phi$ is given by observing that the probability that there are exactly $k=0,1,\ldots,\ell_i$ agents that have their cost greater than $ c^{\star}$ is $\phi_k=1/(\ell_i+1)$. Indeed, {\em a priori}, $\phi_k$ cannot depend on $k$. Therefore $\phi=(b_i+b_i^0+1)\phi_k=(b_i+b_i^0+1)/(\ell +1)$, which, combined with the previous result yields Eq. (\ref{cj}). 

The independence of this result on the distribution of costs of miscoordination implies that strategic uncertainty is relevant {\it even} in the absence of uncertainty on the costs of other players. In what follows, given our emphasis on the collective behavior of the network, we assume $c_j=c$ for all creditors $j$, irrespective of the counterparty. To simplify matters further, we also treat the liquid asset holdings of agents, $b_i^0$ to be constant across the network so that $b_i^0 = b^0$ for all $i$.

\subsection{Network dynamics}
We now present a stylized model that captures the growth and evolution of credit networks by a series of Poisson processes. At rate $\gamma$, each agent $i$ takes out a loan from agent $j$, selected at random from the pool of other investors. This implies $(\ell_i,b_i)\to(\ell_i+1,b_i)$ and $(\ell_j,b_j)\to (\ell_j,b_j+1)$. These loans are unsecured and made without the knowledge of counterparties' current positions.  This may be seen as a reflection of {\it a priori} trust between agents.  The loans mature and are amicably settled by counter-parties with rate $\lambda$. This results in the link removal between agents and an update of their balance sheets.

These two processes would, by themselves, produce a stochastic credit network that belongs to the Erd\"os-R\'enyi random graph ensemble \cite{Renyi1959}, with average degree $\gamma/\lambda$. Therefore, if $\lambda$ is small, i.e., debt is long lived, these two processes produce a dense credit network. 

However, at random Poisson times $t_\nu$, which occur with rate $\nu$, information on the current position $(\ell_i,b_i)$ of agent $i$ is disclosed to all of $i$'s lenders. On the basis of this information, each lenders must decide to either roll-over the loan until maturity or foreclose. The analysis of the foreclosure game provides the lenders with a simple rule to follow, i.e., foreclose their loans if
\begin{equation}
\label{instabcond}
c(\ell_i+1)>b_i+b_0+1\,.
\end{equation}
As a consequence, agent $i$ is said to default and is replaced by a new agent with no links, i.e. $(\ell_i, b_i)\to (0,0)$. Agents, $j$, who previous borrowed from $i$ will each loose one liability, i.e., $(\ell_j,b_j)\to (\ell_j-1,b_j)$. Finally, the lenders, $k$, will loose one asset each, i.e., $(\ell_k,b_k)\to (\ell_k,b_k-1)$. If, instead, Eq. (\ref{instabcond}) is not satisfied, then all of $i$'s counter-parties will rollover their loans. 

Notice that default occurs solely due to the breakdown of trust. We do not assume any exogenous shock that can lead to an agents' failure. This reflects our focus on understanding the role played by trust in credit networks.

\begin{figure}[h]
\begin{center}
\includegraphics[width=13cm]{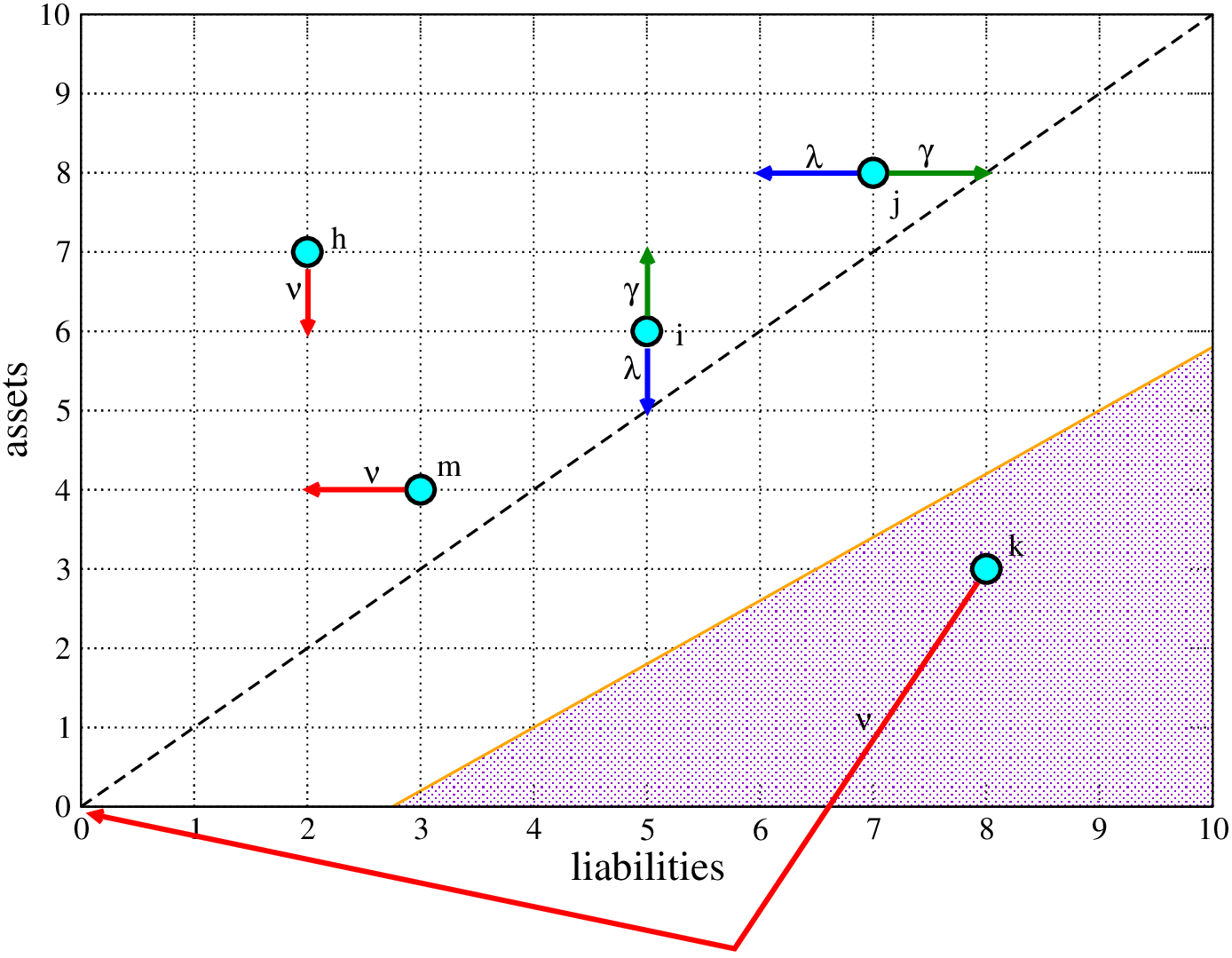}
\end{center}
\caption{Schematic diagram of operations performed in the liabilities-assets, $(\ell,b)$, plane during our network dynamics. The shaded area correspond to where Eq. (\ref{instabcond}) is satisfied and foreclosures take place. With rate $\gamma$,  a credit relationship $i\to j$ is established. Agent $i$ gains an assets ($b_i\to b_i+1$), while $j$ increments the number of liabilities it holds ($\ell_j\to\ell_j+1$). With rate $\lambda$, however this link matures and expires, causing a rearrangement of balance sheets. Finally, with rate $\nu$, debtor $k$ reveals its' balance sheet position, $(\ell_k,b_k)$ to the creditors. If $k$ is found to be in the shaded region, foreclosures take place and $k$ defaults, thereby transporting it back to the origin, i.e., $(\ell_k,b_k)\to (0,0)$. Agent $m$, who had borrowed from $k$, looses one liability ($\ell_m\to\ell_m-1$), while agent $h$ who lent to $k$ looses an asset ($b_h\to b_h-1$).} \label{fig-scheme}
\end{figure}

In summary, an agent's financial state is specified in terms of its position $(\ell_i,b_i)$ in the balance sheet plane. The three processes, (i) link addition at rate $\gamma$ per agent, (ii) link decay at rate $\lambda$ and (iii) information disclosure, at rate $\nu$ per agent, induce a stochastic process in the $(\ell,b)$ plane of balance sheets, as depicted in Fig. \ref{fig-scheme}.

We now turn to the properties of the stationary state of these processes, as a function of the parameters $\gamma,\lambda,\nu,b_0$ and $c$. For simplicity and without loss of generality, we set $\gamma=1$ in what follows, by an appropriate scaling of time.

\section{Results}
\label{sec:results}
\begin{figure}[h]
\begin{center}
\includegraphics[width=13cm]{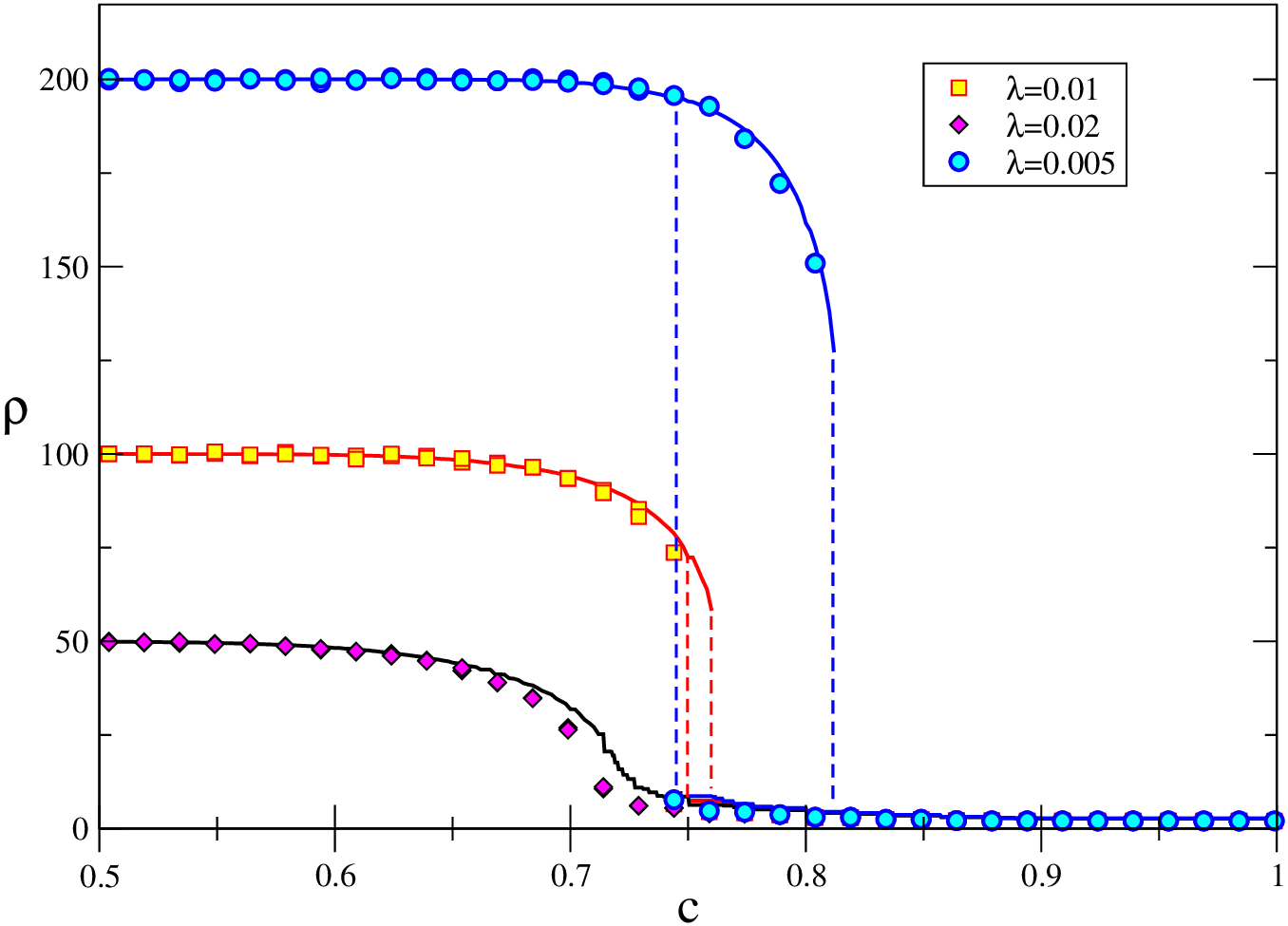}
\end{center}
\caption{Average density $\rho$ in the network as a function of cost $c$ for different values of $\lambda$. The symbols are produced from direct simulations while the lines are from solving the corresponding master equation numerically. In producing the curves we took $\nu\,=\,2.0$ and $N\,=\,1000$.} \label{fig-sim}
\end{figure}

One may probe the collective properties of the stationary state either via direct numerical simulation of the processes or by numerical solving the associated master equation (see Appx. \ref{apx:master} and \cite{Gardiner2009}). In Fig. \ref{fig-sim} we plot simulation results for the average connectivity $\rho$, once the system has reached a stationary state, as a function of the cost of miscoordination, $c$, for different values of debt maturity $\lambda$. We note the following features; (i)  For small $c$, there is a dense network and $\rho =1/\lambda$, indicative of a high level of trust. The news that is released and permeates through the network is encouraging to lenders, who thereby continue to rollover their loans.  (ii) However, for large $c$, lenders perceive that the cost of miscoordination from rolling over their loans is high. Thus, doubts concerning the actions of other lenders leads to the collective foreclose of loans, resulting in a sparse financial network. (iii) For small values of $\lambda$ and in an intermediate range of $c$, we note the coexistence of both dense and sparse network solutions. Finally for larger $\lambda$, one morphs continuously from a dense network to a sparse one, as $c$ is increased.

This hysteresis may be appreciated as a consequence of the self-reinforcing dynamics of trust. Far from the tipping point, a small incremental change in $c$ does not impact the stationary state and we continue to observe the dense network. Once conditions deteriorate -- with increasing $c$ -- beyond the tipping point, a sparse network solution emerges. However, by the same incremental change argument, as one decreases $c$ -- improve conditions -- the sparse network solution is stable and one needs to decrease $c$ to well before the tipping point to regain the dense network solution. This hysteresis is also observed as a function of the liquid assets $b^0$. 

Similar stylized results have been found in other models of network growth \cite{Ehrhardt2006}, which also articulate the underlying mathematical structure. Nevertheless, a qualitative understanding of our results is readily available via a simple approximation of the processes. The key variable is the endogenous rate of link decay, $\mu$, caused by the default of one counterparty. Thus, $\mu$ will crucially depend on the rate $\nu$ at which news is released and the maturity $\lambda$ of debt contracts.

In a dense network, $\mu$ is negligible, whereas in the sparse network it is expected to be sizable. In order to derive an expression for $\mu$, we need to focus on the twin stochastic processes $(\ell^{(t)},b^{(t)})$ for the liability and asset positions for a representative agent evolving in time $t$. This process starts from the origin $(\ell^{(0)},b^{(0)})=(0,0)$ of Fig. \ref{fig-scheme} and drifts toward the top right-hand corner. From any given point on the grid, jumps to the right and up occur at rate $\gamma=1$, whereas jumps to the left or below take place at rate $\lambda+\mu$. In the absence of the absorption process $\nu$, both processes converge to a stationary state, where $\ell^{(t)}$ and $b^{(t)}$ are Poisson variables with mean $1/(\lambda+\mu)$. However, when $\nu$ is ``turned on'', the process is absorbed whenever it is disclosed to be in the shaded region, i.e., $b^{(t)}+b^0+1-c < c\ell^{(t)}$, and restarts at the origin. If $\nu$ is not too large, we may assume that $(\ell^{(t)},b^{(t)})$ attains the stationary state and hence
\begin{equation}
\label{appth}
\mu =  \nu\,{\rm Prob}\left\{ b^{(t)}+b^0+1-c \leq c\ell^{(t)} \right\} \simeq\frac{\nu}{2}{\rm erfc}(Z)\,,
\end{equation}
where ${\rm erfc}(\cdot)$ is the complimentary error function, arising from approximating $b^{(t)}+b^0+1-c - c\ell^{(t)}$ by a Gaussian random variable, and
\begin{equation}
Z = \frac{1-c+b^0(\lambda+\mu)}{\sqrt{2(1+c^2)(\lambda+\mu)}}\,.
\label{Z}
\end{equation}

\begin{figure}[h]
\begin{center}
\includegraphics[width=13cm]{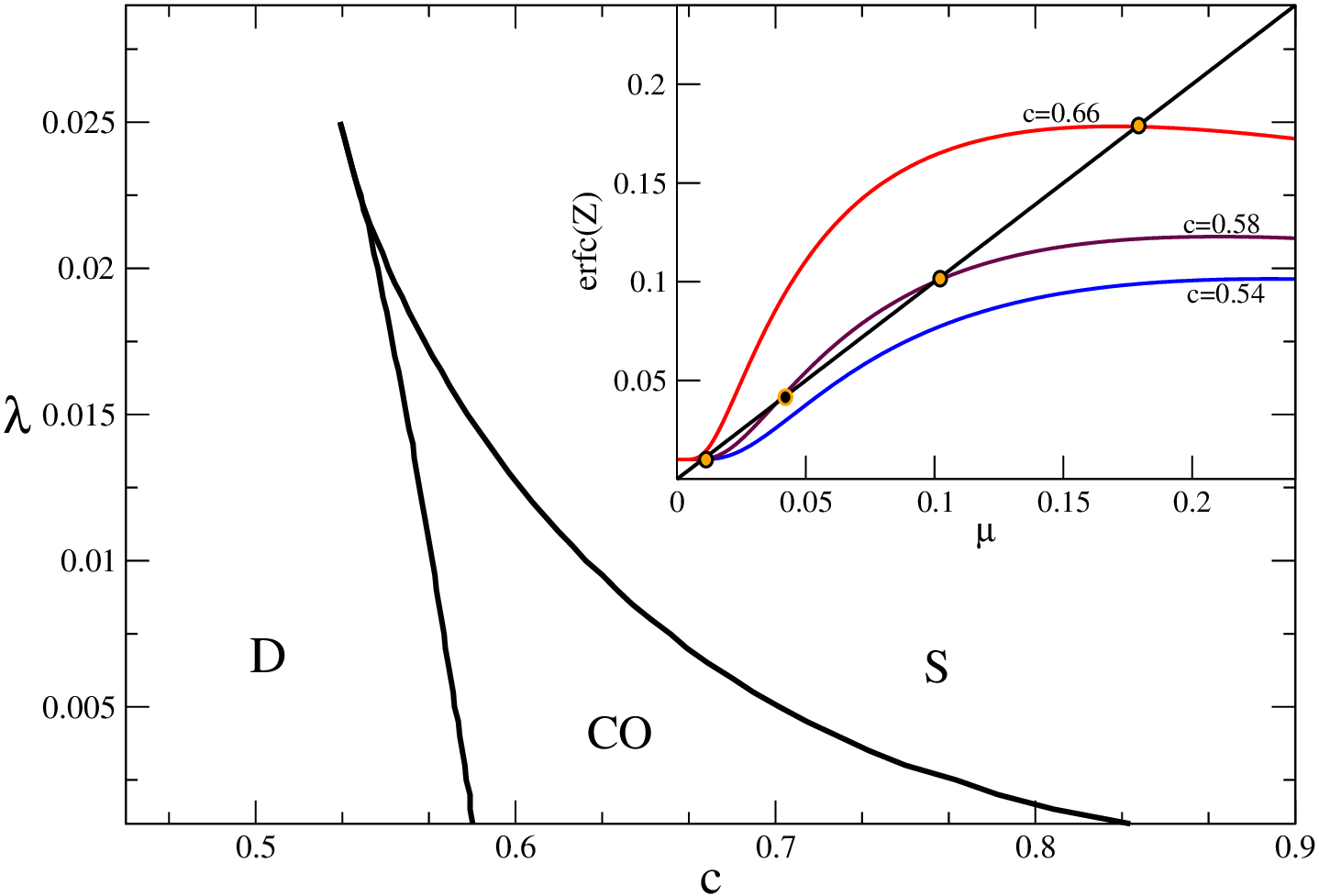}
\end{center}
\caption{ Phase diagram in the $c$ vs. $\lambda$ plane, where the boundaries distinguishes the set of parameters that result in either a dense (D) or sparse (S) network. We also note that for small $\lambda$, there is a third phase of co-existence (CO) between the dense and sparse states. In the Insert we plot ${\rm erfc}(Z)$ as a function of $\mu$, for $\lambda=0.01$, where $Z$ is given by Eq. (\ref{Z}). The different curves correspond to different $c$ values. We not the existence of either one or three fixed points.} \label{fig-phase}
\end{figure}

A graphical solution to Eq. (\ref{appth}) is provided in the inset of Fig. \ref{fig-phase}, where we have either one or three fixed points. Fig. \ref{fig-phase} plots boundaries for regions in the $c$ vs. $\lambda$ plane where these different situations arise. In the dense (D) and sparse (S) phases, one obtains the stable fixed points $\mu\simeq 0$ and $\mu>0$, respectively. In the co-existence (CO) phase, however, the two stable solutions are separated by a third, unstable fixed point. If we impose initial conditions that placed the system to the left of the unstable solution, we would obtain the stable stationary solution $\mu \simeq 0$. Similarly, starting just to the right of the unstable point would yield the solution $\mu>0$. A signature for the transition from ${\rm D}\,\to\,{\rm CO}$ and ${\rm S}\,\to\,{\rm CO}$ is that the slope at the fixed point is exactly one, i.e., it is tangential to the 45 degree line.

Inspection of the argument $Z$ of the ${\rm erfc}$ function provides further insight. For small values of both $c$ and $\lambda$, only one solution with small $\mu\sim e^{-(1-c)/(2\lambda)}$ is possible, as $Z$ is of order $1/\sqrt{\lambda}$. For small $\lambda$ and $c\simeq 1$, instead, the term $1-c$ is negligible with respect to the term $b^0(\lambda+\mu)$. The argument of the ${\rm erfc}$ function is $Z \simeq b^0\sqrt{\lambda+\mu}/2$ and Eq. (\ref{appth}), again, admits one unique solution. In the intermediate range, both solutions are possible, together with a third unstable one.

While precise numerical values of the transition points cannot be accurately obtained, the qualitative features are, however, clear. For example, by increasing $b^0$, or equivalently, decreasing $c$, the curve in the inset of Fig. \ref{fig-phase} moves to the right, thus favoring the dense network phase (low $\mu$). Likewise, decreasing $\nu$ flattens the function, suggesting that the coexistence of solutions is possible only for large values of $\nu$. This is indeed confirmed by numerical simulations. Finally, notice that the dependence on $\lambda$ only enters in the combination $\lambda+\mu$. Hence lowering debt maturity (increasing $\lambda$) is equivalent to shifting the whole curve to the left which again results in the disappearance of the coexistence region, as shown in Fig. \ref{fig-phase} and in the simulations.

\section{Discussion}
\label{sec:discussion}
Our model and results highlight elements that were central to the interbank credit freeze that has characterized the recent global financial crisis.  First, our model shows how the arrival of bad news about a counterparty and the subsequent foreclosure decisions by its creditors can quickly spread across the entire system.  As Fig. \ref{fig-libor} shows, beginning in mid-2007, the financial world was bombarded with jittery news of larger-than-expected and projected losses.  Exogenous disclosure requirements, as modelled by $\nu$, in effect forced banks to release information about their positions, providing investors with signals that helped precipitate the crisis.  In a globalized world, where investors require frequent and better quality information on their investments, it had the effect of making the crisis especially severe and far-reaching.

Second, our model highlights maturity mismatches on balance sheets as a key factor underlying the current crisis.  Banks financed long-term, illiquid, assets (such as special investment vehicles) by short-term borrowing on the interbank market. This situation corresponds to a large ratio $\nu/\lambda$ in our model, where debts have a long maturity compared to the timescale, $1/\nu$, over which banks refinance their debts by convincing creditors to roll over their loans.  It is precisely, and only, in the limit of large $\nu/\lambda$ that a sharp transition such as the one observed in Fig. \ref{fig-sim} can occur.

Third, our model sheds light on the nature of public sector intervention during (and since) the crisis.  The resumption of normality in the interbank markets required a restoration of trust in the balance sheets of key financial institutions.  To facilitate this, central banks cut interest rates to historically low levels, effectively decreasing the cost of miscoordination, $c$.  That these interest rates have been so low and for so long emphasizes the hysteresis entailed in the restoration of trust -- a key feature of our model.  Central banks have also been active in providing emergency lending to troubled institutions as well as sovereign guarantees for borrowing activity, both of which are akin to an increase in $b^0$.

It is worth noting that while the method of analysis suggested by \cite{MorrisShin2003,MorrisShin2008} affords a unique solution to individual coordination games, this uniqueness is lost at the system level; multiple and simultaneous coordination games played on a credit network are characterized by co-existence of equilibria and hysteresis when debt is long-lived

The current policy debate on promoting systemic financial stability has highlighted the importance of liquidity cushions in averting future crises \cite{Caurana2009,Tucker2009}.  Our stylized model lends credence to such considerations. At the system-wide level, our results and numerical simulations indicate that increasing $b^0$, i.e., liquid assets, for all agents (for a given debt maturity $\lambda$) results in dense credit networks with $\rho=1/\lambda$ for larger values of $c$. At the individual level, increasing $b^0_i$ clearly motivates creditor $j$ to roll over loans to agent $i$. 

Relatedly, setting liquidity requirements to be proportional to short-term debt, in the manner of the Greenspan-Guidotti rule \cite{Bussiere1999} for short-term debt-to-reserve ratios in emerging-market countries, can also improve system stability \cite{Squam2009}. In our model, this amounts to setting $b^0_i=\beta_i+\alpha\ell_i$, where $\alpha$ is some pre-defined ratio. From Eq. (\ref{cj}) this is equivalent to reducing the cost of miscoordination $c$ to $c-\alpha$, and replacing $b^0_i$ by $\beta_i-\alpha$.  Clearly, however, the benefits of {\it ex post} regulation of this kind need to be set against the {\it ex ante} costs to banks of such regulation. Requiring banks to hold liquidity cushions may lower the extent of lending {\it ex ante} and the overall implications of such a policy for economic welfare is likely to be unclear.

An alternative to blanket leverage ratios and liquidity requirements is to target such policies on those financial institutions in the network that are most important \cite{Haldane2009}. There are interesting parallels here with the literature on attacks on internet-router networks \cite{Albert2000}.  Extensions of our model along such lines might allow for differential link formation, $\gamma$, or preferential linkage where agents in one sub-group prefer to interact with others in the same sub-group. Agent heterogeneity of this kind holds out the possibility of promising new insights into the design of financial stability policy.

The model presented here is simple. In particular, it does not allow for any macroeconomic variability or the exogenous default by a group (or sub-set) of agents.  Moreover, it seems likely that the parameters of the model will evolve according to economic conditions and will need to be determined endogenously.  For example, in a crisis, agents will strategically disclose information or form links in ways that improve their chances of a public sector bailout.  Incorporating a richer set of economic interactions into a network setting such as ours is an important step for future research.

\begin{appendix}

\section{Foreclosure game}
\label{apx:foreclose}
We now provided a more detailed description and analysis of the foreclosure game, drawing on lines of reasoning provided in \cite{MorrisShin2008}.

\subsection{Basic setup}
There are three distinct time periods, {\it initial}, {\it interim} and {\it final}, which we label $0$, $1$ and $2$, respectively. At the beginning of period $0$, each agent $i\,=\,1,\,\ldots,\,N$ is endowed with one unit of money, which may be used to readily purchase consumption goods during either periods $1$ or $2$. We assume that agents are indifferent between consuming during the interim and final periods. This is formalized by requiring the utility function for each agent to have the simple additive form
\begin{equation}
u(m_1,m_2)\,=\,m_1\,+\,m_2\,,
\end{equation}
where $m_t$ is the consumption during period $t$. This simplification allows us concentrate on measuring coordination in terms of expected returns. 

At the start of period $0$, a pool of investors enter into a {\it short-term loan} agreement with agent $j$, where each loan is for the nominal amount $\delta\,<\,1$. If, for example, creditor $i$ lends to $j$, then during the interim date $i$ is given the choice to either terminate the line of credit (foreclose), or rollover the loan to the final date. If $i$ chooses to rollover its loan and $j$'s investment is successful, then $i$ will receive one unit of money at the end of period $2$.

Thus, at the end of the initial period, $j$ is characterized by three quantities: (i) number of liabilities, $\ell_j\in\mathbb{N}$, i.e., counter-parties that have lent to $j$, (ii) number of illiquid assets, $b_j\,\in\mathbb{N}$, which counts the number of loans given to other agents by $j$ and (iii) level of liquid assets, $b^0_j \geq 0$, which may be thought of as cash reserves. One may interpret the loans made out by $j$ as a result of other agents tendering requests for loans at date $0$ as well.

\subsection{Signals and strategies}
During the interim period, each of $j$'s lenders receive information on the fundamental soundness, $\theta\,>\,0$ of $j$'s investment. If the fundamentals are weak, $j$ would seek an additional injection of capital, which would demand that a large number lenders rollover their loans for the investment to succeed. In particular, if $\ell_j^\prime$ is the number of early foreclosures, then the investment is successful if and only if the number of rollovers satisfies $\ell_j\,-\,\ell_j^\prime\,\geq\, (b_j + b^0_j)$. A high degree of coordination is required by the lenders to see the investment through. In such a case of weak fundamentals, we say $\theta$ is large.

We take $\theta$ is uniformly distributed {\it ex ante} along the positive real axis. The information for agent $i$ is realized as the {\it cost of miscoordination}, $c_i$, from rolling over the loan. In particular,
\begin{equation}
c_i\,=\,\theta\,+\,s_i\,,
\end{equation}
where $s_i\in[-\epsilon,\epsilon]$ is uniformly distributed idiosyncratic noise, with $\epsilon \ll 1$. Thus, while each agent is subject to the same signal generation process, there is asymmetric and imperfect information in the system.

A strategy for agent $i$ is a rule of action that maps each realization of $c_i$ to an action -- whether to rollover the loan of foreclose. We assume that all agents follow a simple switching strategy, i.e., there exists some $c^\star$, such that, if $c_i \geq c^\star$ then $i$ will foreclose its' loan, while roll over if $c_i < c^\star$. On the basis of this strategy, we now solve for the Nash equilibrium of the game.

\subsection{Nash equilibrium}
If lender $i$ forecloses its loan, then the payoff it receives is $-\delta$. However, if $i$ rolls over the loan, then its payoff is
\begin{eqnarray}
      \left\{ \begin{array}{l}
         \,\,\,\,\,\,-\,c_i\,-\,\delta\quad  {\rm if\,}\,\,\,\ell^\prime_j > b_j + b^0_j \\ 
	\\	
	1\,-\,c_i\,-\,\delta\quad {\rm if\,}\,\,\,\ell^\prime_j \leq  b_j + b^0_j\\
         \end{array} \right.\,,
\end{eqnarray}

The argument for the equilibrium solution is as follows. Agent $i$ supposed there is another agent $h$, who has also lent money to $j$. The cost of miscoordination perceived by $h$ is exactly at the switching threshold, i.e., $c_h\,=\,c^\star$. Consequently, $k$ will be indifferent between rolling over and foreclosing its loan. We define $\phi_h$ to be the probability there are sufficiently few foreclosures for the investment to succeed {\it conditional} on $h$'s cost,
\begin{equation}
\label{eq: phi_h}
\phi_h \equiv {\rm Prob}(\ell^\prime_j \leq \ell_j - (b_j + b^0_j)\,|\,c_h = c^\star)\,.
\end{equation}
Next, evaluating the expected payoff, we get the simple relationship
\begin{equation}
\label{eq: expected payoff}
\phi_h\,=\,c^\star\,.
\end{equation}
To evaluate Eq. (\ref{eq: phi_h}), we note that, for given $\theta$, the probability an agent will rollover its loan is the probability its cost lies to the left of $c^\star$, i.e.,
\begin{equation}
\frac{c^\star\,-\,\theta\,+\,\epsilon}{2\,\epsilon}\,.
\end{equation}
Thus, the probability that exactly $n$ out of the $\ell_j$ lenders to $j$ rollover their loans is
\begin{equation}
\psi_n(\theta)\,=\,\left(\begin{array}{c} \ell_j\\ n\end{array}\right )\,\frac{(c^\star\,-\,\theta\,+\,\epsilon)^n (c^\star\,-\,\theta\,+\,\epsilon)^{\ell_j-n}}{(2\epsilon)^{\ell_j}}\,.
\end{equation}
From this we obtain
\begin{eqnarray}
\label{eq:post int}
\phi_h \,=\, \sum_{n\,\geq\,n_j} \int \frac{{\rm d}\theta}{2\epsilon} \psi_n(\theta)\,,
\end{eqnarray}
where $n_j\,=\,\ell_j - (b_j + b^0_j)$. The integral is over the posterior distribution for $\theta$, which is uniform over $[c^\star\,-\,\epsilon,\,c^\star\,+\,\epsilon]$. Note that, despite our prior on $\theta$ was improper, i.e., it had infinite mass, the posterior, which is a conditional distribution is well defined. Finally, by a suitable change of variables, we reduce the integral in Eq. (\ref{eq:post int}) to a Beta function, which we readily evaluate. Combining this result with Eq. (\ref{eq: expected payoff}) we get
\begin{equation}
c^\star\,=\,\frac{b_j + b^0_j + 1}{\ell_j + 1}\,,
\end{equation}
which is independent of $\epsilon$. Having worked out the switching threshold in terms of the cost for a hypothetical agent $h$, agent $i$ adopts this behavior. Moreover, as there was nothing special in our choice of $i$, {\it all} lenders to $j$ will follow a similar exercise and adopt the switching strategy same result.

\section{Master equation}
\label{apx:master}
For a given cost of miscoordination, $c$, the master equation for the joint distribution of liabilities $\ell$ and assets $b$ of a representative agent is given by
\begin{eqnarray}
\partial_t P(\ell,b) & = &  \mu\,\delta_{\ell,0}\,\delta_{b,0} + \gamma\,P(\ell-1,b) + \gamma\,P(\ell,b-1) \nonumber\\
&+& (\lambda+\mu_b)(\ell+1)P(\ell+1,b) \nonumber\\
&+& (\lambda+\mu_l)(b+1)P(\ell,b+1) \nonumber\\
&-& \Big[2\gamma\,+(\lambda+\mu_b)\,\ell +(\lambda+\mu_\ell)\,b\nonumber\\
\label{eq:mastereq} &+&\nu\,\Theta(c(\ell+1)-1-b-b^0)\Big] P(\ell,b)\,,
\end{eqnarray}
where $\partial_t$ is the partial derivative with respect to time and $\Theta(\ldots)$ refers to the Heaviside function. To keep our notation concise, we have left out the time $t$ label in the distribution.

The parameters $\gamma$, $\lambda$ and $\nu$ are exogenously given rates of link creation, dissipation and news arrival, respectively. The rates $\mu$, $\mu_l$ and $\mu_b$ are endogenous default rates, which are self-consistently determined against the stationary distribution of $(\ell,b)$ as
\begin{eqnarray}
\mu & = & \nu \sum_{\ell,b}\Theta(c(\ell+1)1-b-b^0)P(\ell,b)\,,\\
\mu_\ell & = & \frac{\nu}{\langle\ell\rangle} \sum_{\ell,b}\Theta(c(\ell+1)1-b-b^0))\,\ell\, P(\ell,b)\,,\\
\label{eq:barnu_b}\mu_b & = & \frac{\nu}{\langle b\rangle} \sum_{\ell,b}\Theta(c(\ell+1)1-b-b^0))\,b\,P(\ell,b)\,.
\end{eqnarray}
Here, $\langle b \rangle$ and $\langle \ell \rangle$ are the mean assets and liabilities, respectively. The angled brackets refers to the average over $P(\ell, b)$, which in fact yields $\langle b \rangle\,=\,\langle \ell \rangle$.

We can understand the master equation via simple geometric considerations. If we consider any arbitrary point $(\ell,b)$ in the interior of our lattice, the probability that an agent has this balance sheet positions is given by the probability there will be an incremental {\it hop} to $(\ell,b)$ from one of the neighboring sites. The rate at which there will be a hop from the left or bottom site is the rate $\gamma$ at which either a new liability or asset, respectively, is added. The rate at which a hop occurs from either above or the right is simply the rate at which an asset or liability are lost. In the latter case, this rate may be decomposed into two aspect: (i) the natural dissipation of a link, i.e., $\lambda$ and (ii) the probability that our representative agent had borrowed from another agent who defaulted and lost all links after revealing, with rate $\nu$, its balance sheet to its creditors. The rate at which such incidents occur is $\mu_b$. A similar argument may be used to construct the rate at which assets are lost as well. Whenever an agent defaults, it is stripped of all its asset and replaced by a new agent, who starts at $(\ell,b)\,=\,(0,0)$. The first term in Equation (\ref{eq:mastereq}) reflects this action. 

A full analytical solution for Equation (\ref{eq:mastereq})-(\ref{eq:barnu_b}) is complicated due to the presence of various non-linearities. We can nevertheless numerically solve the system for the stationary state.
\end{appendix}

\end{document}